\documentclass[12pt,letterpaper]{article}
\pdfoutput=1

\usepackage{graphicx}

\usepackage[utf8]{inputenc}

\usepackage{color}
\usepackage{slashed}
\usepackage{color}
\definecolor{refkey}{rgb}{1,0,0}
\definecolor{labelkey}{rgb}{0,0,1}
\usepackage{bbold,amsfonts,amssymb}
\usepackage{physics,amsfonts,amssymb,amsmath}
\usepackage{MnSymbol,wasysym}
\usepackage{esint}
\usepackage{rotating}

\usepackage{etoolbox}

\usepackage{graphicx}
\usepackage{caption}
\usepackage{subcaption}
\usepackage{booktabs}

\usepackage{listings}

\usepackage{hyperref}

\numberwithin{equation}{section}
\newcommand{\be}{\begin{equation}}
\newcommand{\ee}{\end{equation}}
\newcommand{\ben}{\begin{displaymath}}
\newcommand{\een}{\end{displaymath}}
\newcommand{\bea}{\begin{eqnarray}}
\newcommand{\eea}{\end{eqnarray}}
\newcommand{\bean}{\begin{eqnarray*}}
\newcommand{\eean}{\end{eqnarray*}}

\def\a {\alpha}

\def\e {\epsilon}

\newcommand{\eg}{{\it e.g.}}
\newcommand{\ie}{{\it i.e.}}

\newcommand{\commentout}[1]{}

\usepackage{amsmath}

\renewcommand{\theequation}{\arabic{section}.\arabic{equation}}

\newcommand{\beq}{\begin{equation}}
\newcommand{\eeq}{\end{equation}}
\newcommand{\beqr}{\begin{displaymath}}
\newcommand{\eeqr}{\end{displaymath}}
\newcommand{\beqa}{\begin{eqnarray}}
\newcommand{\eeqa}{\end{eqnarray}}
\newcommand{\beqar}{\begin{eqnarray*}}
\newcommand{\eeqar}{\end{eqnarray*}}
\newcommand{\cN}{{\cal N}}

\newcommand{\cO}{{\cal O}}

\newcommand{\non}{\nonumber}

\newcommand{\sign}{\ensuremath{\mathrm{sign}}}

\begin{document}
%%%%%%%%%% TITLEPAGE %%%%%%%%%%%%

\title{\Large \bf Non-analyticity in Holographic Complexity near Critical points}

\author{
	Uday Sood$^1$,
	Martin Kruczenski$^{1,2}$\thanks{E-mail: \texttt{usood@purdue.edu, markru@purdue.edu.}} \\
	$^1$ Dep. of Physics and Astronomy,	and \\ $^2$ Purdue Quantum Science and Engineering Institute  \\
	 Purdue University, W. Lafayette, IN, USA. }

\maketitle

\begin{abstract}
 The region near a critical point is studied using holographic models of second-order phase transitions. In a previous paper, we argued that the quantum circuit complexity of the vacuum ($C_0$) is the largest at the critical point. When deforming away from the critical point by a term $\int d^d x \, \tau \, \cO_\Delta$ the complexity $C(\tau)$ has a piece non-analytic in $\tau$, namely 
 $C_0 -C(\tau) \sim |\tau-\tau_c|^{\nu(d-1)} + \mathrm{analytic} $. Here, as usual, $\nu=\frac{1}{d-\Delta}$ and $\xi$ is the correlation length $\xi\sim |\tau-\tau_c|^{-\nu}$ and there are possible logarithmic corrections to this expression.  
 That was derived using numerical results for the Bose-Hubbard model and general scaling considerations. In this paper, we show that the same is valid in the case of holographic complexity providing evidence that the results are universal, and at the same time providing evidence for holographic computations of complexity.
	
\end{abstract}

\clearpage

\tableofcontents
\newpage

\section{Introduction}
In a previous work \cite{Sood:2021cuz}, we studied the circuit complexity of the ground state of systems near quantum critical points using numerical and field-theoretical methods. The notion of complexity used in that work followed from identifying optimal circuits with geodesics on the space of circuits similar to \cite{Nielsen2006AGA, nielsen2006quantum, Dowling2008TheGO} in the context of quantum computing and \cite{jefferson2017circuit, Chapman:2017rqy, Khan:2018rzm, Hackl:2018ptj, Guo:2018kzl, Chapman:2018hou, Bhattacharyya:2018bbv, Jiang:2018nzg} in the context of quantum field theory. Related calculations of complexity in many body systems near quantum phase transitions can be found in the recent works \cite{Afrasiar:2022efk, Pal:2022ptv} for the LMG (Lipkin-Meshkov-Glick) model, \cite{Meng:2021wmz} for the Proca theory and the closely related work \cite{Huang:2021xjl} on the N-site Bose-Hubbard model.

In this work, we continue our study of complexity this time through the lens of known holographic conjectures. The first such conjecture is that the complexity should be dual to the volume of the extremal codimension-one bulk hypersurface which meets the asymptotic boundary on the time slice where the boundary state is defined \cite{Stanford:2014jda}. The second conjecture states that complexity should be dual to the gravitational action evaluated on the WDW patch of the spacetime \cite{Brown:2015bva, Brown:2015lvg}. The WDW patch is the region of spacetime enclosed by past and future light sheets sent into the bulk from the time slice on the boundary. These proposals have led to numerous insights, for ex. see  \cite{Stanford:2014jda, Carmi:2017jqz, Susskind:2020gnl, Susskind:2020wwe, Hashemi:2019aop}.
Using these conjectures, we study the critical behavior of the complexity in holographic RG flows \cite{Skenderis:2002wp, Bianchi:2001de}. \\
A nice feature of the holographic correspondence is that it allows us to study RG flows of strongly coupled field theories using weakly coupled dual gravitational theories. The geometries that will be of interest to us are asymptotically AdS geometries. The interpretation in the field theory is that of a perturbed CFT which undergoes a renormalization group flow. In the holographic correspondence, the radial coordinate of the AdS geometry is interpreted in terms of the energy scale in the field theory, therefore a dependence of a bulk field on the radial coordinate represents an RG flow. On the field theory side, perturbations are introduced by adding a source term to the CFT Lagrangian or by giving a vacuum expectation value to a certain operator. These two types of perturbations correspond to the non-normalizable and normalizable modes of the dual bulk fields, respectively \cite{PhysRevD.59.104021}.  \\
In our previous study, we found that the complexity is the largest at the critical point $\tau=\tau_c$ and, as $\tau \rightarrow \tau_c$, has a non-analytic piece that behaves as $|\tau - \tau_c|^{\nu (d-1)}$ for a d-dimensional spacetime field theory. We also found that this term is independent of the UV cutoff. We find all these features in both the volume and the action calculations near holographic critical points. This is the main result of this paper. In addition, we found that the analytic terms in $C(\tau)$ were all regularization-dependent and hence ambiguous even after subtracting $C_0$. This continues to hold in the case of holographic complexity. \\
The organization of the paper is as follows: we start with an explanation of the general gravitational setup that we use for RG flow geometries in Section \ref{bulk gravity}. We include expressions for the general form that the volume and action calculations take for these geometries. In Section \ref{non-analytic terms}, we show that all such expressions contain a term that is in general non-analytic in the deformation coupling $\tau$ by deriving a scheme to extract such a term which we denote by $v_0$ and $i_0$ for the volume and action respectively. Section \ref{n=1 flow} looks at an example where the volume and action can be computed analytically. This is the case of the $\cN = 1$ flow geometry. Next, in Section \ref{ambiguity}, we show that the complexity defined using the volume or action conjectures has various ambiguities as it is regulator-dependent. We look at a few such ambiguities in the volume and action. We find that the non-analytic term discussed previously is universal and independent of the cutoff. This is similar to the nature of complexity found in \cite{Sood:2021cuz} where we found a non-universal complexity but a universal non-analytic term in the field theory calculation for the complexity near the $O(2)$ fixed point in a Gaussian approximation. Finally, we summarize our results with the discussion in Section \ref{conclusions}.

%----------------------------------------------------------------------------------	-----------------------------------------------------------------------	-------------------------------------------

\section{Bulk Gravity Setup}
\label{bulk gravity}
We want to study holographic complexity near a critical point that defines a UV CFT dual to AdS. Upon deforming it by an IR-relevant operator the theory may flow in the IR to another CFT or a gapped phase. In the dual gravitational theory, we consider an asymptotically AdS background of the form 
\begin{align}
ds^2 = \frac{L^2}{z^2} (\eta_{\mu\nu} dx^{\mu} dx^{\nu} + \frac{dz^2}{f(z)})    
\label{general metric}
\end{align}
We require that the function $f(z)$ have a scale $\xi$ and be of the form $f(z/\xi)$ such that $f(z)\simeq 1$ when $z\ll \xi$. Then, $\xi$ defines a correlation length such that for length scales smaller than $\xi$ the theory is described by a CFT. For larger length scales we can have a flow to another CFT or to a gapped phase based on how $f(z)$ behaves in the region $z>\xi$. The boundary deformation by an IR-relevant operator $\cO$ is modeled by turning on a bulk scalar field $\Phi$ with a mass related to the conformal dimension of $\cO$ in the bulk. Consider then the bulk action    
\begin{align}
I_{bulk} &=\frac{1}{16\pi G_N} \int d^{D}x  \sqrt{-g} \left( R-\frac{1}{2} g^{AB} \partial_A \Phi \partial_B \Phi - V(\Phi) \right)
\label{gravity action}
\end{align}
where $g_{AB}$ is the spacetime metric and $V$ is the scalar potential which has a critical point at $\Phi =0$, satisfying $V(0) = -\frac{d(d-1)}{L^2}$. \\
When we have $f(z) \rightarrow \frac{L^2}{L^2_{0}} > 1$ as $z \rightarrow \infty$, the geometry again becomes AdS with radius $L_0$. For consistency, the potential again has a critical point at the $z \rightarrow \infty$ value of the scalars $V_0 = -\frac{d(d-1)}{L_0^2}$. Such geometries are known as domain wall geometries in the literature. \cite{Girardello:1999bd, Freedman:1999gp}\\
On the other hand, flows to a gapped phase involve an $f(z)$ that keeps on growing in the interior and typically blows up with some positive power of $z$ as
$z \rightarrow \infty$ \cite{Girardello:1998pd}. The spacetime is singular here but typically becomes non-singular when considering extra dimensions and generically the space-time caps off at a finite proper distance from the boundary.\\
In addition to how the metric behaves in the interior, the RG flow geometries are also distinguished based on whether the leading deformation is source-like or vev-like. The leading behavior in the scalar dual to the relevant deformation ($\Delta <  d$) of the field theory which we call $\Phi$ determines whether we have a source-like deformation 
\begin{align}
\Phi = \Phi_{(s)} z^{d-\Delta} + ....
\end{align}
or a vev-like deformation
\begin{align}
\Phi = \Phi_{(v)} z^{\Delta} + ....
\end{align}
in the standard quantization framework. \\
The equations of motion following from the action in Eq. $\ref{gravity action}$ are 
\begin{align}
\label{equations of motion}
R_{AB} &= \frac{1}{2}\partial_A \Phi \partial_B \Phi + \frac{1}{d-1}g_{AB} V(\Phi)  \\
\frac{1}{\sqrt{-g}} &\partial_A (\sqrt{-g}g^{AB}\partial_B\Phi) - \frac{\delta V}{\delta \Phi} = 0
\end{align}
Near the boundary, the equation for $R_{zz} + R_{tt}$ gives,
\begin{align}
zf'(z) = \frac{(d-\Delta)^2 \Phi_{(s)}^2}{d-1}z^{2(d-\Delta)}
\end{align}
Thus the leading correction to $f(z)$ is positive.
\begin{align}
f(z) = 1 + \frac{(d-\Delta)\Phi_{s}^2}{2(d-1)} z^{2(d-\Delta)} + ...
\end{align}
The critical exponent $\nu$ determines the scaling of $\Phi_{(s)}$ with the correlation length in the dual field theory \footnote{$\nu$ is defined in the standard way via $\xi \sim \Phi_{(s)}^{-\nu}$. Since $\int d^{d}x \Phi_{(s)} O_{\Delta}$ is a term in the action, $\frac{1}{\nu} = d-\Delta$.}. Thus, the above equation is of the form $f(z) = 1 + (z/\xi)^{2(d-\Delta)} + ...$ \footnote{If $\Delta$ = d/2, then the leading order term in f(z) is $(z/\xi)^{d} (\log(z/\xi))^2$. This also follows from a near boundary analysis of Einstein equations and the fact that even the asymptotic leading order behavior of the scalar has a $z^{d/2}\log(z)$ term.}. A similar near-boundary analysis can be done in the case of vev deformations which yields an asymptotic form of $f(z)$ given by 
\begin{align}
f(z) = 1+ \frac{\Delta}{2(d-1)}z^{2\Delta}\, \Phi_{(v)}^2 + ...
\end{align}
The positivity of the leading correction to $f(z)$ near the boundary follows more generally from the null energy theorem and Einstein equations which imply that $f(z)$ is monotonically increasing $\ie$ $\partial_z f > 0$. \cite{Freedman:1999gp, Liu_2013}.
\commentout{When there is a single scalar, the leading contribution apart from the cosmological constant term to the stress-energy tensor 
\begin{align}
T_{AB} &= -\frac{2}{\sqrt{-g}}\frac{\delta I_{scalar}}{\delta g^{AB}}  
\end{align}
near the boundary comes from the derivatives of the $\Phi_{(s)}$ or $\Phi_{(v)}$ terms along with the mass term in the potential. Using the mass-dimension relation $m^2L^2 = \Delta (\Delta-d)$, we have
\begin{align}
T_{zz} &= \frac{d}{4\kappa^2} (d-\Delta) \Phi_{(s)}^2 z^{2(d-1-\Delta)} + ... \\
T_{\mu \nu} &= \frac{1}{2\kappa^2}\eta_{\mu \nu} (d-\Delta) (\Delta - \frac{d}{2}) \Phi_{(s)}^{2} z^{2(d-1-\Delta)} + ...
\end{align}
for the source deformations. The quantity $T_{tt} + T_{zz}$ is manifestly positive. This can be explicitly checked for this case but it is also a consequence of the null energy condition applied near the boundary. This implies that the leading correction to $f(z)$, in this case, is positive.
\begin{align}
f(z) = 1 + c_d z^{2(d-\Delta)} + ...
\end{align}
with $c_d > 0$. In fact, $c_d \propto \Phi_{(s)}^2$ with the proportionality constant depending on d and $\Delta$. }

\begin{figure}[h]		
\centering
\begin{subfigure}{0.4\textwidth}
\centering
\includegraphics[width = \textwidth]{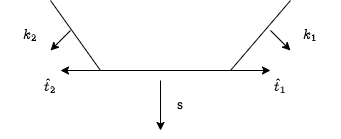}
\caption{Timelike-null joints}
\label{timelike-null joint fig}
\end{subfigure}
\hfill
\begin{subfigure}{0.4\textwidth}
\centering
\includegraphics[width = \textwidth]{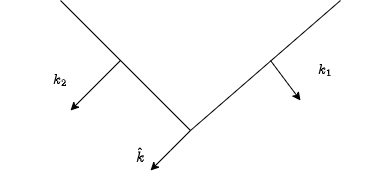}
\caption{Null-null joints}
\label{null-null joint fig}
\end{subfigure}
\caption{Joints in AdS WdW spacetime}
\end{figure}

The maximal-volume computation is straightforward for the class of metrics in eq. $\ref{general metric}$. The maximal volume slices are fixed-$t$ slices and the maximal volume is invariant under boundary time translations. Thus we can fix the boundary time at which to evaluate the maximal volume to be $t=0$. Then, the maximal slice satisfies $t=0$ for all $z$. The volume of such a slice is 
\begin{align}
V_{\Sigma}[f] = \sigma_{d-1} L^d \int_{z_0}^{Z_0} dz \frac{1}{z^d \sqrt{f(z)}}
\end{align}
The volume of these slices would be infinite for $z_0 \rightarrow 0$ because proper distances near $z=0$ diverge and these slices extend to $z=0$. These UV divergences are expected in the field theory definitions of complexity as well \cite{Sood:2021cuz, jefferson2017circuit}. So, we use a regulated version of AdS space with a cutoff surface at $z=z_0$. Also, $\sigma_{d-1}$ is the volume of the spatial field theory background which is finite after making the spatial coordinates of the boundary theory periodic. In principle, such periodic identification is singular at the Poincare horizon $z \rightarrow \infty$ which can be avoided by using a large-z cutoff $Z_0$. However, this is unnecessary here since the complexity we compute is not singular in the limit $Z_0 \rightarrow \infty$. It would be useful to note that for $d\geq2$ a fixed-$t$ AdS slice has a volume given by
\begin{align}
V_{\Sigma}(AdS_{d+1}) = \frac{\sigma_{d-1}L^d}{(d-1)\epsilon^{d-1}}
\end{align} 
where $z_0 = \epsilon$. \\
We also compute the action of the scalar-gravity system on the WDW patch of spacetimes given by the class of metrics in eq. $\ref{general metric}$. This is yet another coordinate-invariant object in the bulk. The full action on the WDW requires an accounting of boundary terms as well along with $I_{bulk}$ \cite{Chapman:2016hwi}. These terms come from co-dimension 1 boundary segments as well as co-dimension 2 joints formed by the intersection of these segments. Any spacelike/timelike segments require the addition of the Gibbons-Hawking-York term \cite{York:1972sj, Gibbons:1976ue} and any joints formed by these surfaces require additional terms \cite{Hayward:1993my, Brill:1994mb} for the variational principle to be well defined. Similarly, null segments require a boundary term \cite{Parattu:2015gga, Lehner:2016vdi} and so do the joints formed by the intersection of null segments with spacelike/timelike boundary segments. 
\begin{figure}[h]
\centering
\includegraphics[width=0.5\textwidth]{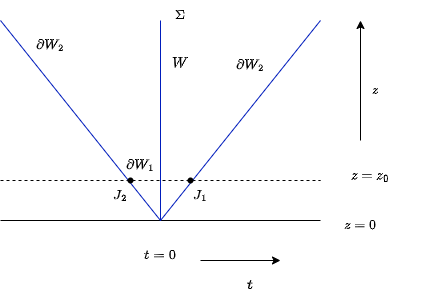}
\caption{WDW region for AdS spacetime}
\label{ads_wdw}
\end{figure}
\begin{align}
I = I_{bulk}(W) &+ I_{GHY}(\partial W_1) + I_{null}(\partial W_2) + I_{jnts} (J)
\end{align}
In the above equation, $\partial W_1$ refers to any spacelike/timelike boundary segments of the WDW patch, $\partial W_2$ to null boundary segments, and $J$ to any joints formed between boundary segments. Naively, it may seem that the definition of the WDW patch should imply that one only has null boundary segments. However, this may not be the case as some regularization schemes may introduce non-null surfaces Fig, \ref{ads_wdw}. \\
The precise formula for $I_{bulk}$ involves evaluating eq. $\ref{gravity action}$ for the specific choice of functions $f$ and $\Phi$ on the regulated WDW patch. The second term is the usual Gibbons-Hawking-York term for spacelike/timelike boundaries $\partial W_1$. It is 
\begin{align}
 I_{GHY}(\partial W_1) = \frac{1}{8 \pi G_N} \int d^{d}x \sqrt{|h|} K
\end{align}
with $h$ being the induced metric on $\partial W_1$ and K being the trace of the extrinsic curvature. \footnote{All normals are taken to point outwards w.r.t. W} The third term is the contribution from null boundaries $\partial W_2$. It is given by 
\begin{align}
I_{null}(\partial W_2) = -\frac{1}{8 \pi G_N} \int d\lambda d^{d-1}\theta \sqrt{|\gamma|} \kappa
\end{align}
This term is required for the variational principle to be well-defined whenever a spacetime has null boundaries. This piece depends on the parametrization \footnote{ The variation and the equations of motion do not.} through $\kappa$ which measures the non-affinity in $\lambda$. $\gamma$ is the $(d-1)-$dimensional metric on the $\theta$ coordinates. A choice of $\lambda$ gives a null normal $k^A$ to the surface which satisfies
\begin{align}
k^A \grad_A k_B = \kappa k_B
\end{align}
A choice of $\lambda$ can set this term to zero and we make this choice in this section and the next. In Sec \ref{ambiguity}, we consider a different choice and check whether that changes this term in the action. Finally, the term evaluated on joints is 
\begin{align}
I_{jnts}(J) = \frac{1}{8 \pi G_N} \int d^{d-1}x \sqrt{\sigma} a
\end{align}
where for timelike-null joints $a = -\sign(k.s)\sign(k.\hat{t}) \log |k.s|$ where k is the null normal and s is the one form spacelike unit normal to the timelike surface. $\hat{t}$ is a tangent vector in the tangent space of the timelike surface, orthogonal to the junction and again pointing outwards as shown in fig. \ref{timelike-null joint fig}.

We compute the action for $AdS_{D}$ as a warmup. Any scalars are turned off in this case and therefore we only have the cosmological constant term from the scalar. For the $AdS_{D}$ space, we have $R = - \frac{d(d+1)}{L^2}$ which gives 
\begin{align}
I_{bulk} (AdS_{D}) = -\frac{dL^{d-1}}{\kappa^2}\int_{WDW}d^{D}x \frac{1}{z^{d+1}}
\end{align}
In this case, the WDW patch is enclosed by $t = \pm z$ light sheets. Including regulators at small $z=\epsilon$ and large $z=Z_0$, we have 
\begin{align}
I_{bulk} (AdS_{D}) = -\frac{d\sigma_{d-1} L^{d-1}}{(d-1)4\pi G_N}\left( \frac{1}{\epsilon^{d-1}} - \frac{1}{Z_0^{d-1}}  \right)
\end{align}
We find that $I_{bulk} (AdS_{D})$ is negative and that we can take the $Z_0 \rightarrow \infty$ limit. \commentout{ giving specifically for $AdS_5$
\begin{align}
I_{bulk} (AdS_5) = -\frac{\sigma_3 L^{3}}{3\pi G_N}\frac{1}{\epsilon^{3}} 
\end{align}}
For the GHY term, we find that the surface at $z=Z_0$ again gives a vanishing contribution as $Z_0 \rightarrow \infty$. For general $d \geq 2$, one obtains
\begin{align}
I_{GHY}(AdS_{D}) = \frac{d\sigma_{d-1}L^{d-1}}{4 \pi G_N} \left( \frac{1}{\epsilon^{d-1}} - \frac{1}{Z_0^{d-1}} \right)
\end{align}
\commentout{For the $AdS_5$ case with $Z_0 \rightarrow \infty$, this contribution is $\sigma_3L^3/\pi G_N \epsilon^3$.} As mentioned above, we can affinely parametrize the null generators and make any contributions from the null boundaries vanish. The WDW patch for vacuum AdS has four joints all of which are null-timelike joints. The outward null normals corresponding to affine parametrizations are $dt - dz$ and $-dt-dz$ respectively. The joint contributions give 
\begin{align}
I_{jnts}(AdS_D) = -\frac{L^{d-1}\sigma_{d-1}}{4 \pi G_N}\left(  \frac{\log{\epsilon/L}}{\epsilon^{d-1}} - \frac{\log{Z_0/L}}{Z_0^{d-1}} \right)
\end{align}
\commentout{to get $-L^3\sigma_3 \log{(\epsilon/L)}/4 \pi G_N\epsilon^3$ for $AdS_5$.}
We can again drop the $Z_0$ terms. Adding up all these contributions, the full action for $AdS_{d+1}$ is 
\begin{align}
I(AdS_{d+1}) = \frac{\sigma_{d-1} L^{d-1}}{4 \pi G_N\epsilon^{d-1}} \left(  \frac{d(d-2)}{d-1} + \log{L/\epsilon}  \right)
\end{align}
We find that the total action is positive here, even though the bulk contribution was negative. Hence, we find that the boundary terms are important for the action to be a valid measure of complexity.
\commentout{
\begin{align}
I(AdS_5) = \frac{\sigma_3 L^3}{4 \pi G_N\epsilon^3} \left(  \frac{8}{3} + \log{L/\epsilon}  \right)
\end{align}
}

%---------------------------------------------------------------------------------------------------------------------------------------------------------------------------------------------------------------------------------------------------------------------------------------------------------
\section{Non-analyticity near the critical point}
\label{non-analytic terms}
This section shows that near the critical point, the volume and action have a piece non-analytic in the deformation coupling $\tau$ and we derive the general form for this piece. \\
The volume of interest is $V_{\Sigma}[f]$. To isolate the non-analytic piece, we use the variables $u = z/\xi$ and also take $Z_0 \rightarrow \infty$ as we did in the previous section.
\begin{align}
V_{\Sigma}[f] = \frac{\sigma_{d-1} L^d}{\xi^{d-1}} \int_{z_0/\xi}^{\infty} du \frac{1}{u^d \sqrt{f(u)}}
\end{align}
where, as discussed later in sec.\ref{ambiguity}, we introduce a value for $z_0$ defined through
\beq
 \epsilon = \xi \int_0^{z_0/\xi}\!\!\! \frac{du}{\sqrt{f(u)}} \label{cutoff}
\eeq
with $\epsilon $ a fixed UV cut-off. The function $f(u)$ is increasing with $u$ and $f(0)=1$. 
We start with some general considerations to show that the complexity from the volume prescription decreases as we move away from the critical point, \ie\ $d V_\Sigma[f]/d\tau<0$ for $\tau>0$. Indeed
\beq
 \frac{d V_\Sigma[f]}{d\xi} = \sigma_{d-1} L^d \left\{ -(d-1) \frac{1}{\xi^{d}}\int_{z_0/\xi}^{\infty} du \frac{1}{u^d \sqrt{f(u)}} + \frac{\epsilon}{z_0^d\xi} \right\}  \label{derV}
\eeq
where we used, from eq. \ref{cutoff}, that
\beq
 \frac{d}{d\xi} \left(\frac{z_0}{\xi}\right) = -\frac{\epsilon}{\xi^2}\sqrt{f(z_0/\xi)}
\eeq
Eq. \ref{derV} can be rewritten as 
\beq
 \frac{d V_\Sigma[f]}{d\xi} = \sigma_{d-1} L^d \frac{(d-1)}{\xi^d} \int_{z_0/\xi}^\infty du\, \left(\frac{\epsilon}{z_0}- \frac{1}{\sqrt{f(u)}} \right)
\eeq
Now, due to the fact that $f(u)$ is increasing with $u$ we obtain from eq.\ref{cutoff} 
\beq
 \frac{\epsilon}{z_0} > \frac{1}{\sqrt{f(z_0/\xi)}} > \frac{1}{\sqrt{f(u)}}, \ \ \ \mbox{if} \ \ \ u>\frac{z_0}{\xi}
\eeq
implying that $ \frac{d V_\Sigma[f]}{d\xi}>0$, and, in view of $\xi\sim \tau^{-\nu}$ with $\nu>0$:
\beq
 \frac{d V_\Sigma[f]}{d\tau} < 0
\eeq
showing that the complexity from the volume prescription indeed has a peak at the transition $\tau=0$. To show that this peak has a non-analytic part we start by recalling that near the boundary $z \ll \xi$, 
\begin{align}
f(z) = 1 +  \sum_{m=2}^{\infty} c_m \left(\frac{z}{\xi}\right)^{m \alpha} +  \cO\left((z/\xi)^{2\Delta}\right)
\label{powerseriesf}
\end{align}
where $\alpha = d- \Delta$ for a source deformation and $c_2 = 1$ \cite{Hung:2011ta,Liu_2013}. We consider only source-like deformations here but the calculation in this section can be generalized to the vev case as well. To see a power series solution for the scalar-gravity action of this form, see Appendix \ref{power series solution}. The second series of terms come from the subleading scalar terms discussed in the previous section. Next, we introduce a parameter $\delta$ which is small $\delta << 1$, and use it to break up the integral.
\begin{align}
V_{\Sigma}[f] = \frac{\sigma_{d-1} L^d}{\xi^{d-1}} \left[ \int_{z_0/\xi}^{\delta} du \frac{1}{u^d \sqrt{f(u)}} + \int_{\delta}^{\infty} du \frac{1}{u^d \sqrt{f(u)}} \right]
\end{align}
Since $\delta$ is small, we can expand the function $f$ using the small $u$ expansion in the first integral.
\begin{align}
f(u)^{-1/2} = 1  - \frac{1}{2} \sum_{m=2}^{\infty} \tilde{c}_m u^{m \alpha} +  \cO\left(u^{2\Delta}\right)
\end{align}
where $\tilde{c}_2=1$ and since we are interested in separating the divergences, the last $\cO(u^{2\Delta})$ term can be ignored because it gives a finite contribution to the integral because $2\Delta>d$. After performing the first integral and replacing $\xi=\tau^{-\nu}$, we obtain
\begin{align}
V_{\Sigma}[f] = \sigma_{d-1} L^d \left[ \frac{1}{(d-1)z_0^{d-1}} + \frac{1}{2} \sum_{m=2}^{\infty} \tilde{c}_m \frac{z_0^{m\alpha - d + 1}}{m\alpha - d + 1} \tau^m  + \ldots + \tau^{\nu(d-1)}v_0 \right] 
\label{C1a}
\end{align}
with $v_0$ given by
\begin{align}
v_0 =\lim_{\delta \rightarrow 0} \left( \int^{\infty}_{\delta} du \frac{1}{u^d \sqrt{f(u)}} - \frac{1}{(d-1)\delta^{d-1}} - \frac{1}{2}\sum_{m=2}^{\infty} \tilde{c}_m \frac{\delta^{m\alpha - d + 1}}{m\alpha - d + 1} + \ldots \right)
\label{v0}
\end{align}
This limit is finite since we subtracted all the infinite pieces. The ellipsis ($\ldots$) in eq. $\ref{C1a}$ refer to other analytic terms similar to $\tau^{m}$ and in eq. $\ref{v0}$ refer to other divergent pieces in the limit $\delta \rightarrow 0$. 
The result \eqref{C1a} shows that the divergent part is analytic in $\tau$ but there is a non-analytic contribution proportional to $\tau^{\nu(d-1)}$ independent of the cut-off $z_0$. Notice that, if for some integer $m_0$ we have $m_0 \alpha = (d-1)$ then the complexity will have a (non-analytic) logarithmic term $\ln \xi$. On the other hand, we will have $\tau^{\nu(d-1)}=t^{2m_0}$, and that contribution will be analytic but still independent of the cut-off. For some toy model calculations of $v_0$, see Appendix \ref{toy model}.

A similar analysis gives the non-analytic piece $i_0$ for the CA prescription of the complexity \footnote{The action has a purely gravitational part considered here and a part coming from the scalar. The sum of both should have a peak at $\tau=0$ as we show later in particular examples.}. This term comes from the bulk part of the action. Since the gravitational terms are always present in the bulk action, we restrict to these terms here and show that they contain a piece non-analytic in $\tau$.
\begin{align}
I_{bulk} &=\frac{1}{16\pi G_N} \int d^{D}x  \sqrt{-g} \left( R - V(0) \right)  \\
&= \frac{d L^{d-1}\sigma_{d-1}}{8\pi G_N} \int_{z_0}^{\infty} \frac{dz}{z^{d+1}\sqrt{f(z)}} \big( d-1 + z f'(z) - (d+1)f(z)\big) \int_{0}^{z} \frac{dy}{\sqrt{f(y)}}   \non
\end{align}
Changing variables to $u = z/\xi$ and $w = y/\xi$ so that we have
\begin{align}
I_{bulk} = \frac{dL^{d-1}\sigma_{d-1}}{8\pi G_N \xi^{d-1}} \int_{z_0/\xi}^{\infty} \frac{du}{u^{d+1}\sqrt{f(u)}}\big( d - 1 + u f'(u) - (d+1)f(u) \big) \int_{0}^{u}\frac{dw}{\sqrt{f(w)}}
\end{align}
By again introducing a scale $\delta$ between $z_0/\xi$ and $\infty$ with $\delta << 1$ as before,
\begin{align}
I_{bulk} = - \frac{dL^{d-1}\sigma_{d-1}}{8\pi G_N} &\left( \frac{2}{(d-1)z_0^{d-1}} + \sum_{m=2}^{\infty} f_m \tau^m z_0^{m\alpha - d + 1} \right) + L^{d-1}\sigma_{d-1}\tau^{\nu(d-1)}i_0  
\end{align}
with $i_0$ given by
\begin{align}
i_0 = \frac{d}{8 \pi G_N}\lim_{\delta \rightarrow 0} &\left( \int_{\delta}^{\infty} \frac{du}{u^{d+1}\sqrt{f(u)}} \big( d - 1 + u f'(u) - (d+1)f(u)  \big) \int_{0}^{u} \frac{dw}{\sqrt{f(w)}} \right. \\ \non
&\left. + \frac{2}{(d-1)\delta^{d-1}} + \sum_{m=2}^{\infty}f_m\delta^{m\alpha-d+1} \right)
\end{align}

\commentout{\begin{align}
I_{bulk} = - \frac{dL^{d-1}\sigma_{d-1}}{8\pi G_N} &\left( \frac{2}{(d-1)z_0^{d-1}} + \sum_{n=0}^{\infty}\frac{f_n (1+n(1+n+d))}{(n+1)(n+d+1)}z_0^{n+1}       \right.     \\  \non
&\left. + \sum_{m=2}^{\infty} \frac{(m\alpha - d)(m\alpha + 1) +  1}{(m\alpha + 1)(m \alpha - d + 1)} \tau^m z_0^{m\alpha - d + 1} + \ldots \right) \\ \non
&+ L^{d-1}\sigma_{d-1}\tau^{\nu(d-1)}i_0 
\end{align}
with $i_0$ given by
\begin{align}
i_0 = \frac{d}{8 \pi G_N}\lim_{\delta \rightarrow 0} &\left( \int_{\delta}^{\infty} \frac{du}{u^{d+1}\sqrt{f(u)}} \big( d - 1 + u f'(u) - (d+1)f(u)  \big) \int_{0}^{u} \frac{dw}{\sqrt{f(w)}} \right. \\ \non
&\left. + \frac{2}{(d-1)\delta^{d-1}} + \sum_{m=2}^{\infty}\frac{(m\alpha - d)(m\alpha + 1) + 1}{(m\alpha+1)(m\alpha-d+1)}\delta^{m\alpha-d+1}  + \ldots \right)
\end{align}}
Here, the coefficients $f_m$ can be obtained from the coefficients $c_m$ and, the $I_{GHY}, I_{null}, I_{J}$ parts of the gravitational action do not give any contributions to $i_0$ since they are analytic. We show this in Appendix \ref{non-analytic boundary calculation}.

%-----------------------------------------------

\section{$\cal N$ $=1$ Gapped Flow}
\label{n=1 flow}
The metric is known analytically for the flow of $\mathcal{N}=4$ SYM theory to a confining theory under a mass-like source deformation with dimension $\Delta = 3$ in the UV \cite{Girardello:1998pd}. The metric is asymptotically $AdS_{5}$ with $f(z)$ and $\Phi(z)$ given by \footnote{Note that $\xi$ differs here from definition in previous sections by a constant factor of $\sqrt{2}$ for ease of notation.}
% gppz paper
\begin{align}
&f(z) = (1+z^2/\xi^2)^2 \\
&\Phi(z) = \frac{\sqrt{3}}{2} \log{\frac{\sqrt{\xi^2+z^2}+z}{\sqrt{\xi^2+z^2}-z}}
\end{align}
The potential $V(\Phi)$ in the action for the scalar is 
\begin{align}
V(\Phi) = -\frac{3}{2L^2} ( 3 + 4 \cosh{(2\Phi/\sqrt{3})} +  \cosh{(2\Phi/\sqrt{3})}^2  )
\end{align}

The near boundary asymptotics $z << \xi$ for the scalar and metric are consistent with a source deformation with $\nu = 1$.
\begin{align}
&f(z) = 1+\frac{2z^2}{\xi^2} +... \label{fPhi} \\
&\Phi(z) = \frac{\sqrt{3}z}{\xi} + ...
\end{align}
 Note here that since $\nu (d-1) = 3$ takes a special value i.e an integer, therefore $\tau^{\nu(d-1)}=\tau^3$ is analytic in $\tau$. However, this term is still distinguished because it is the only term in the series with an odd power of $\tau$. In that sense it is natural to use a variable $\tilde{\tau}=\sqrt{\tau}$ in which case $\tilde{\tau}^{\frac{3}{2}}$ would be the non-analytic term near $\tau=0$.   \\
Of course, the $\mathcal{N} = 1$ geometry just becomes the AdS geometry when $\xi \rightarrow \infty$ or when the scalar source is turned off. Therefore, we can expand the difference in maximal volumes or actions in powers $z_0/\xi$. 
$V_{\Sigma}$ can be computed analytically and is given by
\begin{align}
 \frac{V_{\Sigma}(\xi)}{\sigma_3L^4} &= \frac{1}{3z_0^3} - \frac{1}{\xi^2 z_0}
 + \frac{\pi}{2\xi^3} - \frac{1}{\xi^3} \tan^{-1}{\frac{z_0}{\xi}} . 
\end{align}
Following up on the discussion above eq.\ref{fPhi}, the function on the right-hand side is even under $\xi\rightarrow -\xi$ except for the term $\frac{\pi}{2\xi^3}$ that leads to the $\tau^3$ term that we interpret as the universal non-analytic term in this case. 
For $z_0/\xi$ small, replacing $\xi=\tau^{-1}$ and writing explicitly the terms that do not vanish as $z_0\rightarrow 0$ we obtain
\begin{align}
  \frac{V_{\Sigma}(\tau)}{\sigma_3L^4}  &=\frac{1}{3z_0^3} - \frac{\tau^2}{z_0}
 + \frac{\pi}{2}\, \tau^3 + \cO(z_0)
 \end{align}
We find regulator-dependent leading terms, a regulator-independent sub-leading term in $V_{\Sigma}$ that contains the only odd power of $\tau$, and then terms that go to 0 as $z_0 \rightarrow 0$ and that only contain even powers of $\tau$. \\
For the WDW action in the $\mathcal{N}=1$ flow case, the potential evaluated on the solution of the scalar is 
\begin{align}
V = -\frac{6}{L^2} \left( 2+3u^2+u^4  \right)
\end{align}
with $u = z/\xi$. As in the vacuum AdS case, we have to regulate the geometry with a small z cutoff $z_0$ and a large z cutoff at $Z_0$ to regulate the singularity at $z \rightarrow \infty$. To label the WDW region, it is useful to define light-cone coordinates $v_1$ and $v_2$ using
\begin{align}	
dv_1 &= dt-dz/\sqrt{f}  
\label{dv_1}   \\
dv_2 &= -(dt+dz/\sqrt{f} )
\label{dv_2}
\end{align}
Integrating, one has $v_1 = t - \xi \tan^{-1} z/\xi$ and $v_2 = - (t + \xi \tan^{-1} z/\xi)$. Then the region enclosed by $v_1=0$ and $v_2=0$ in the regulated geometry is the WDW patch. These coordinates are also useful as $\ref{dv_1}$, $\ref{dv_2}$ gives null normals $k_1 = k_{1\mu}dx^{\mu} = dv_1$ and $k_2 =  k_{2\mu}dx^{\mu} = dv_2$ that satisfy the geodesic equation with $\kappa = 0$ and point outwards. In terms of the familiar Poincare coordinates, the WDW patch is bounded between the rays $t=\pm \xi \arctan{z/\xi}$ on the $t-z$ plane. For this spacetime,
\begin{align}
I_{bulk} = \frac{\sigma_3L^3}{8\pi G_N\xi^3} \int_{z_0/\xi}^{Z_0/\xi} du \frac{\tan^{-1}{u}}{u^5}\left(  \frac{u^2}{2} - 8 \right)
\end{align}
We can again take the $Z_0 \rightarrow \infty$ in this integral without encountering any difficulties. The integral gives 
\begin{align}
I_{bulk} = \frac{\sigma_3L^3}{8\pi G_N \xi^3} \left( -\frac{2}{3u_0^3} - \frac{1}{2u_0^4}(1-\frac{u_0^2}{8}-\frac{9u_0^2}{2})\tan^{-1}u_0 + \frac{9}{4u_0}-\frac{9\pi}{8}  \right)
\end{align}
%\begin{align}
%I_{bulk} = \frac{\sigma_3 L^3}{8 \pi G_N \xi^3} \left(  -\frac{2}{3u_0^3}-\frac{1}{u_0^4}\left(1-\frac{u_0^2}{8}\right)(\pi-2\cot^{-1}u_0)  + \frac{9}{4u_0}\left(1 - u_0\cot^{-1}u_0 \right)\right)
%\end{align}
with $u_0 = z_0/\xi$. The terms in the bracket can be expanded for small $u_0$ to get the leading behavior in the deformation away from the critical point,
\begin{align}
I_{bulk}(\tau) = \frac{\sigma_3L^3}{8 \pi G_N} \left( -\frac{8}{3z_0^3} + \frac{19}{6 z_0}\, \tau^2 - \frac{9\pi}{8}\, \tau^3 + \cO(z_0) \right)   \non
\end{align}
The first term in the expression is the only one that survives when $\tau \rightarrow 0$ and is the familiar AdS contribution. The sub-leading universal piece again goes like $\tau^3$, the only odd power of $\tau$ in the expansion. The coefficients of this term are different for the two different prescriptions which we called $v_0$ and $i_0$ in Sec \ref{non-analytic terms}.\\
We next compute the GHY term for the timelike surfaces at $z=z_0$ and $z=Z_0$. These give 
\begin{align}
I_{GHY} = \frac{L^3\sigma_3}{\pi G_N}\frac{\xi \tan^{-1}{z_0/\xi}}{z_0^4}\left( 1+z_0^2/\xi^2 \right)
\end{align}
as the $z=Z_0$ gives zero as we take $Z_0$ to the Poincare horizon. For small $u_0$  
\begin{align}
I_{GHY}(\tau) = \frac{L^3\sigma_3}{\pi G_N} \left( \frac{1}{z_0^3} + \frac{2}{3 z_0}\, \tau^2 + \cO(z_0) \right)
\end{align}
Here we find as expected that the $GHY$ piece does not contribute to the universal piece in the WDW action. Next, we compute the contributions from the joints in the WDW region. As in the AdS case, there are four joints each of them formed by null-timelike intersections. The two joints connected to the $z=Z_0$ do not contribute while the $z=z_0$ do contribute
\begin{align}
I_{joints} = -\frac{\sigma_3L^3}{4 \pi G_N z_0^3}\log{z_0/L}
\end{align}
This quantity also does not contribute to the universal term in the action.
Adding up all the action contributions, we get
\begin{align}
I(\tau) = \frac{\sigma_3 L^3}{8 \pi G_N} \left(  \frac{1}{3z_0^3}(16+6\log{L/z_0}) +  \frac{17}{2z_0}\, \tau^2 - \frac{9\pi}{8}\, \tau^3  + \cO(z_0) \right)
\end{align}
Here again, the total action is positive as in the AdS case.
Thus, the coefficients $v_0$ and $i_0$ for this flow are
\begin{align}
v_0 &= \frac{\pi}{2}   \\
i_0 &= -\frac{9}{64 G_N}
\end{align}

\section{Holographic Complexity and Ambiguities}
\label{ambiguity}
In general, the holographic complexity defined using either the CV or the CA prescriptions is an ambiguous quantity even after choosing one prescription (see \eg\ \cite{Chapman:2016hwi}). However, in this section, by considering the various types of ambiguities we show that the non-analytic term calculated in the previous sections is free from such ambiguities and therefore continues to be meaningful. Still, there is a difference in the coefficient of the universal term as computed by the two prescriptions, and how that matches different definitions of complexity in the field theory remains to be seen.  \\
The quantities calculated in the previous section $V_{\Sigma}$ and $I$ have been interpreted as representing the complexity of the quantum state dual to the geometry according to 
\begin{align}
C_V &= \frac{V_{\Sigma}}{G_N l}  
\end{align}
for CV and 
\begin{align}
C_A &= \frac{I}{\pi \hbar}  
\end{align}
for the CA conjecture \footnote{The reference state wrt which these quantities should be considered as complexities and the length scale $l$ in the CV prescription are both ambiguous. We take $l=L$ in this paper.}. \\
In this way, the previous sections allow us to compute $C_V$ and $C_A$ for the $\cN=1$ flow geometry. We see that both $C_V$ and $C_A$ for this geometry start with the AdS contributions and then have corrections coming from the deformation $\tau$. So we define the subtracted quantities $\delta V_{\Sigma}$ and $\delta I$ for the $\mathcal{N} = 1$ geometry by looking at the difference
\begin{align}
\delta V_{\Sigma} &= V_{\Sigma} (AdS_{5}) - V_{\Sigma} (\mathcal{N} = 1)  \\
\delta I &=  I(AdS_5) - I(\cN = 1)
\end{align}
These are the analogs of the complexity of formation for black hole spacetimes where the deformation is in the temperature. In earlier holographic calculations, this quantity was found to be free of several ambiguities that may arise in the calculation of the volume or the action \cite{Chapman:2016hwi}. 
In this section, we are comparing the action and volumes in two different asymptotically AdS geometries. We use a systematic way of applying the cutoff in the Fefferman-Graham coordinates $y = \epsilon$ where the metric is of the form
\begin{align}
ds^2 = \frac{L^2}{y^2}(dy^2 + g_{\mu\nu}(x,y)dx^{\mu}dx^{\nu})
\end{align}
The quantity $z_0(\epsilon)$ then has an expansion for general $f(z)$ of the form \ref{powerseriesf}
\begin{align}
z_0 (\epsilon) = \epsilon \big( 1  + \sum_{m=2}^{\infty} \frac{d_m }{2(m\alpha+1)}\big(\frac{\epsilon}{\xi}\big)^{m\alpha} + \cO\left((\epsilon/\xi)^{2\Delta}\right)  \big)
\end{align}
with $d_2 = 1$. We find that $z_0$ is analytic in $\tau$ up to the $\cO\left((\epsilon/\xi)^{2\Delta}\right)$ terms in the above expansion. 
%Another useful result is that this also holds for the quantity $\sum_{n} z_0^{n\alpha}$.\\ 
The small-z cutoff in the case of the $\cN=1$ flow is 
\begin{align}
z_0(\epsilon) = \frac{\epsilon}{\sqrt{1-\tau^2 \epsilon^2}}
\end{align}
The pure AdS cutoff and the $\mathcal{N} = 1$ flow cutoff differ by
\begin{align}
z_0(\epsilon) - \epsilon = \frac{\tau^2 \epsilon^3}{2} + O(\epsilon^{5})
\end{align}
Using this result, we write down the expressions for $\delta C_V$ and $\delta C_A$ 
\begin{align}
\delta C_V = \frac{\sigma_3 L^3}{G_N} \left( \frac{3 \tau^2}{2 \epsilon} - \frac{\pi \tau^3}{2} \right)
\end{align}
and 
\begin{align}
\delta C_A = \frac{\sigma_3L^3}{16\pi^2G_N \hbar} \left(  \frac{\tau^2}{\epsilon}(1+ 6\log{L/\epsilon}) + \frac{9\pi \tau^3}{4} \right)
\end{align}
Here in both the prescriptions, we find that $\delta C > 0$ for $\tau > 0$ $\ie$ the complexity is the largest at the critical point and decreases as we move away. We showed that this holds generally in sec. \ref{non-analytic terms} for the volume case, but here we find that this also true for the action prescription in this particular example. 

\subsection{Choice of cutoff}
We find that the subtracted quantity $\delta C$ defined using either the CV or the CA proposal is still dependent on the choice of the cutoff $\epsilon$. This is the first ambiguity that we encounter.

\subsection{Null-null joint at the cutoff surfaces}
Instead of regulating the patch as we did in the previous section, one could regulate the patch in such a way that the null normals meet at the $z = z_0$ surface as in Fig \ref{null-null joint fig}. Then instead of two null-timelike joints, we have one null-null joint. This procedure also shifts the null normals infinitesimally so that they are now,
\begin{align}
t = \xi \tan^{-1} z/\xi - \xi \tan^{-1} z_0/\xi   \\
t =  \xi \tan^{-1} z_0/\xi - \xi \tan^{-1} z/\xi 
\end{align}
for the $\mathcal{N} = 1$ case and
\begin{align}
 t = z -\epsilon  \\
 t = \epsilon - z 
\end{align}
for the AdS case. Due to this infinitesimal shift, $I_{bulk}$ changes. With this choice, the new $I_{bulk}$ expressions are
\begin{align}
I_{bulk} (AdS_5) &= - \frac{\sigma_3 L^3}{12 \pi G_N \epsilon^3}    \\
I_{bulk}(\mathcal{N} = 1) &= I_{bulk} (AdS_5) + \frac{\sigma_3 L^3}{8 \pi G_N} \left( \frac{13}{4\epsilon}\, \tau^2 - \frac{9 \pi}{8}\, \tau^3   \right)
\end{align}
up to terms that vanish when $\epsilon \rightarrow 0$. In this new scheme, the Gibbons-Hawking-York terms are 0 for both geometries because the timelike surface at the cutoff is no longer present. The null piece can also be taken to be zero in both cases by choosing $\kappa = 0$. The character of the joint terms changes as we now have a null-null joint. For this joint, $a =- \sign(k.\tilde{k})\sign(\hat{k}.\tilde{k}) \log{(k.\tilde{k}/2)}$ where $\hat{k}$ is in the tangent space of the boundary region which has the normal $k$ $(\hat{k}.k = 0)$, null and pointing outwards and away from the joint while $\tilde{k}$ is the normal of the other null region. With this prescription, the AdS joint term is 
\begin{align}
I_{J} (AdS_5) = -\frac{\sigma_3 L^3}{4 \pi G_N \epsilon^3} \log{\epsilon /L}
\end{align}
and the $\mathcal{N} = 1$ term is 
\begin{align}
I_{J} (\mathcal{N} = 1) = -\frac{\sigma_3 L^3}{4 \pi G_N z_0^3} \log{z_0 /L}
\end{align}
which is the same as the joint term in the case with joints as in Fig. \ref{timelike-null joint fig}. Thus, the difference $\delta I_{J}$ is the same. However, $\delta I_{bulk}$ and $\delta I_{GHY}$ are different. The quantity $\delta C_A$ still continues to be positive and is given by
\begin{align}
\delta C_A = \frac{3\sigma_3 L^3}{16 \pi^2 G_N \hbar}  \left(  \frac{\tau^2}{\epsilon}(2\log{L/\epsilon} - 3/2) + \frac{3\pi}{4}\, \tau^3 + \cO(\epsilon) \right)
\end{align}

\subsection{Null-normal normalization and parametrization}
Another ambiguity that can be kept track of is in the normalization of the null normals. We introduce $c > 0$ so that $k_1.\hat{t} = c$ instead of 1. In this case, 
\begin{align}
I_{J} (AdS_5) &= -\frac{\sigma_3 L^3}{4 \pi G_N \epsilon^3} \log{(c \e/L)}   \\
I_{J} (\mathcal{N} = 1) &= -\frac{\sigma_3 L^3}{4 \pi G_N z_0(\epsilon)^3} \log{(c z_0(\epsilon)/L)}   \\
\delta I_{J}  &= \frac{\sigma_3 L^3 \tau^2}{8 \pi G_N \epsilon} \left( 1 + 3\log{L/c \epsilon} \right)
\end{align}
We find that $\delta I$ and thus $\delta C_A$ has some $c$ dependence in the leading term. \\
Next, we consider a different parametrization $\tilde{\lambda} (\lambda)$ from the previously chosen affine case $\lambda$ for the null surfaces. This induces new normals $\tilde{k_1}$ and $\tilde{k_2}$ and hence we have to recalculate both the null piece and the joint piece in the action. For simplicity, we consider whether $\delta I$ changes when $\kappa$ changes from zero to a constant not equal to zero. A different $\delta C_A$ would suggest that the complexity has another ambiguity that comes from the parametrization chosen for the boundary regions and therefore the existing definition requires an additional choice of parametrization to be specified. An affine parametrization $\lambda$ satisfies 
\begin{align}
d\lambda = -\frac{L^2}{z^2}\frac{dz}{\sqrt{f}}
\label{affine}
\end{align}
This gives the null normals $k_1$ and $k_2$ discussed earlier that satisfy $\kappa = 0$. We want to make a change $\lambda \rightarrow \tilde{\lambda} (\lambda)$ such that $\tilde{\kappa}$ is a constant different from zero.
We consider a change of parametrization of the form 
\begin{align}
\frac{d\tilde{\lambda}}{d\lambda} = e^{-\beta}
\end{align}
Then, the new normals $\tilde{k}^A$ satisfy $\tilde{k}^A = e^{\beta} k^{A}$. Then $\tilde{\kappa} = \frac{d}{d\lambda} e^{\beta}$ or equivalently $\tilde{\kappa} = \frac{d}{d\tilde{\lambda}} \beta$ so that $e^{\beta} = 1 + \tilde{\kappa} \lambda$. The quantity in the integrand $\tilde{\kappa} d\tilde{\lambda} = d\beta = \frac{\tilde{\kappa}d\lambda}{1+\tilde{\kappa}\lambda}$ can be written as an integral over $z$ using eq. $\ref{affine}$. For the AdS case,
\begin{align}
e^{\beta} = 1 + \frac{\tilde{\kappa}L^2}{z}
\end{align}
 while for the $\mathcal{N} = 1$ case, 
 \begin{align}
e^{\beta} = 1 + \frac{\tilde{\kappa}L^2}{z} +  \frac{\tilde{\kappa}L^2}{\xi} \tan^{-1} z/\xi
\end{align}
The integrals for $I_{null}$ are
\begin{align}
I_{null}(AdS_5) &= \frac{\tilde{\kappa}\sigma_3 L^5}{4 \pi G_N} \int_{\e}^{\infty} \frac{dz}{z^4(z+\tilde{\kappa}L^2)}   \\
I_{null} (\mathcal{N} = 1) &=  \frac{\tilde{\kappa}\sigma_3 L^5 \xi^2}{4 \pi G_N}  \int_{z_0}^{\infty} \frac{dz}{z^5(z^2 + \xi^2)(1+\tilde{\kappa}L^2/z + (\tilde{\kappa}L^2/\xi )\tan^{-1} z/\xi)}
\end{align}
Using the notation $\tilde{\kappa} L^2 = k$, the AdS integral is given by
\begin{align}
I_{null}(AdS_5) = \frac{\sigma_3 L^3}{4 \pi G_N} \left( \frac{1}{3\epsilon^3} - \frac{1}{2k\epsilon^2} + \frac{1}{k^2 \epsilon} + \frac{1}{k^3} \log{\epsilon/k}   \right) + \cO(\epsilon)
\end{align}
Keeping terms upto $\tau^5$, the $\cN=1$ integral gives 
\begin{align}
I_{null}(\cN = 1) = \frac{\sigma_3 L^3}{4 \pi G_N} \left( A_0 + A_2 \tau^2 + A_4 \tau^4 + A_5 \tau^5 + \cO(\tau^6) \right)  
\end{align}
with
\begin{align}
A_0 &= \frac{1}{3 \epsilon^3} - \frac{1}{2 k \epsilon^2} + \frac{1}{k^2 \epsilon} + \frac{1}{k^3} \log{\epsilon/k}   \\
A_2 &=  -  \frac{5}{2\epsilon} - \frac{1}{2k} - \frac{3}{k} \log{\epsilon/k}  \\
A_4 &= \frac{k}{6} \left( -11 + 6 \log{t k} \right)  \\
A_5 &= \frac{k^2}{12} \left( -22 + \pi + 6 \pi \log{2}   \right)
\end{align}	
where we again drop terms that go to $0$ as $\epsilon \rightarrow 0$.
\commentout{
\begin{align}
I_{null}(AdS_5) &= \frac{\sigma_3 L^3}{4\pi G_N z_0^3} B[-z_0/\epsilon;4,0]  \\
I_{null}(\cN=1) &= \frac{\sigma_3 L^3\epsilon}{4\pi G_N\xi^4} \int_{z_0(\epsilon)/\xi}^{\infty} \frac{du}{u^4(1+u^2)\left(u(1+\frac{\epsilon}{\xi}\tan^{-1}u) + \frac{\epsilon}{\xi}\right)} \\
& \approx \frac{\sigma_3 L^3\epsilon}{4\pi G_N\xi^4} \int_{z_0(\epsilon)/\xi}^{\infty} \frac{du}{u^4(1+u^2)\left(u + \frac{\epsilon}{\xi}\right)} \\ \non
&=\frac{\sigma_3L^3}{4\pi G_N z_0^3(1+z_0^2/\xi^2)}\left( -\frac{z_0^2}{6\xi^2} + \frac{1}{2}(\Psi(5/2) - \Psi(2)))   \right)
\end{align}
where $B[a;b,c]$ is the incomplete beta function and $\Psi(z)$ is the digamma function. \\}

The sign in the new $I_{J}$ is unchanged since $e^{\beta}$ is positive definite, so only the term in the $\log$ is modified. The modification is 
\begin{align}
I_{J} (AdS_5) &= -\frac{\sigma_3 L^3}{4 \pi G_N \epsilon^3} \log{((\e + k)/L)}   \\
I_{J} (\mathcal{N} = 1) &= -\frac{\sigma_3 L^3}{4 \pi G_N z_0^3} \log{((z_0 + k + \tau z_0 k  \tan^{-1}\tau z_0)/L)}
\end{align}
so that 
\begin{align}
\delta I_{null} &= \frac{\sigma_3 L^3}{4\pi G_N} \left( A_2 \tau^2 + A_4 \tau^4 + A_5 \tau^5 + \cO(\tau^6)  \right) \\
\delta I_{J} &=  \frac{3\sigma_3 L^3 \tau^2}{16 \pi G_N \epsilon} \left(  1+2\log{L/2\epsilon} \right)
\end{align}

 \subsection{Functional redefinitions of the null surface}
Finally, we have functional redefinitions of the null surface. We look at shifts in the quantity $a$ by constant $a_0$ at all the joints. 
\begin{align}
&I_{J} (AdS_5) = \frac{\sigma_3 L^3}{4 \pi G_N \epsilon^3}(a_0 - \log{\e/L } )  \\
&I_{J} (\mathcal{N} = 1) = \frac{\sigma_3 L^3}{4 \pi G_N z_0^3} (a_0 - \log{z_0/L})   \\
&\delta I_{J} = \frac{\sigma_3 L^3 \tau^2}{8 \pi G_N \epsilon} (1+3(a_0 - \log{\e/L})) 
\end{align}

\subsection{Scalar boundary term}
Following \cite{Bernamonti:2020bcf}, we consider the case when the scalar action is modified by an extra term
\begin{align}
I_{s}(\partial W_1) = \frac{1}{16\pi G_N} \int d^{d}x  \sqrt{|h|} \frac{1}{2} \Phi s^{A} \partial_A \Phi
\end{align}
Here, the unit normal $s$ and the region $\partial W_1$ are defined in sec. \ref{bulk gravity}. We find that just as in the case of boundary terms for the gravitational action, this term is analytic in $\tau$ for the $N = 1$ case and does not affect the universal piece of the action complexity calculation. Since it is zero for the pure AdS case when $\tau =0$, the quantity $\delta I_{s} (\partial W_1)$ is given by 
\begin{align}
\delta I_{s} (\partial W_1) = \frac{3 \sigma_3 L^3}{16 \pi G_N} \left( \frac{\tau^2}{\epsilon} + \cO(\epsilon) \right)
\end{align}
\\
Thus we find that the term that goes like $\xi^{-3}$ or $\tau^3$ is independent of $\epsilon$ and therefore universal $\ie$ it does not change under the ambiguities discussed above, while the $\epsilon$-dependent terms are not universal. Analogous quantities to $C_V$ and $C_A$ can also be computed for situations when the temperature T is the only deformation away from the critical point. $\delta V$ or $\delta I$ is then proportional to the complexity of formation for the thermal state based on whether we use the CV or the CA proposal. For the black hole geometry dual to the thermal state, one finds that in both these cases, the quantity $\delta C(T) = C(T) - C(T=0)$ is independent of the regularization $\epsilon$. Not only is the complexity of formation free from the ambiguity of being regulator-dependent, but it is also free from other ambiguities that we looked at above \cite{Chapman:2016hwi}. Therefore, these complexities of formation $\delta C$ appear to be meaningful in this case but we find that this does not carry over to this RG flow example.\\

\subsection{Some comments on ambiguity in the field-theoretic complexity}
Thus we see that any holographic dual to the complexity should also accommodate such ambiguities in RG flow computations of complexity in field theories. Such a possibility may exist as it can be shown that the quantity $\delta C$ can depend on the cutoff in existing definitions of complexity for discretized Gaussian field theories. As an example, consider $\kappa$-complexities for an $h$-dimensional spatial lattice \cite{jefferson2017circuit}
\beq
 C_\kappa = \frac{1}{2^{\kappa-1}} \sum_p \left|\ln\left(\frac{\sqrt{p^2+m^2}}{\omega_0}\right)\right|^\kappa = \frac{1}{2^{\kappa-1}}\frac{V\Omega_{h-1}}{(2\pi)^{h}} \int_0^\Lambda p^{h-1} dp \left|\ln\left(\frac{\sqrt{p^2+m^2}}{\omega_0}\right)\right|^\kappa
\eeq 
where $V$ is the volume of the system and $\Omega_{h-1}= \frac{2\pi^{h/2}}{\Gamma(h/2)}$ is the volume of the unit sphere $S^{h-1}$ with momentum cut-off $\Lambda=\sqrt{\omega_0^2-m^2}$ where the logarithm vanishes. For $\omega_0\gg m$ we can take $\omega_0\sim \frac{1}{a}$ where $a$ is the lattice spacing. In the quantum field theory, we take $\omega_0$ as a UV cut-off. We get
\beq
 c_\kappa = \frac{C_\kappa}{V} = \frac{1}{2^{\kappa-1}}\frac{\Omega_{h-1}}{(2\pi)^{h}} \int_0^{\sqrt{\omega_0^2-m^2}} p^{h-1} dp \left|\ln\left(\frac{\sqrt{p^2+m^2}}{\omega_0}\right)\right|^\kappa
\eeq
A direct computation of $c_1$ gives
\beqa
c_1 = \frac{\Omega_{h-1}}{(2\pi)^h m^2}\frac{(\omega_0^2-m^2)^{1+h/2}}{h(h+2)} {}_2 F_{1}[1,1+h/2,2+h/2,1-\frac{\omega_0^2}{m^2}]
\eeqa
For $h=2$:
\beqa
 c^{h=2}_1 &=& \frac{1}{8\pi} \left\{m^2\ln\frac{m^2}{\omega_0^2}-m^2+\omega_0^2\right\} \\
 c^{h=2}_2 &=& \frac{1}{16\pi} \left\{-\frac{m^2}{2}\ln^2\frac{m^2}{\omega_0^2} + m^2\ln\frac{m^2}{\omega_0^2}-m^2+\omega_0^2\right\}
\eeqa
This suggests that the quantity $\delta c = c(m=0) - c(m)$ still depends on the UV cutoff. For example,
\beq
\delta c_1^{h=2} = \frac{1}{4\pi} \log{\omega_0/m} + \frac{m^2}{8\pi}
\eeq
Thus here again in the context of a free theory near a Gaussian fixed point $(m = 0)$, we find that $\delta c$ exhibits regulator dependence. How the other ambiguities should be accommodated or interpreted on the field theory side is still unclear. However, once the cost function or $\kappa$ and the reference state are fixed, the term $\tau^{\nu h}$ is again independent of the cutoff and therefore universal in that sense.

\commentout{In this way, the previous sections allow us to compute $C_V$ and $C_A$ for the $\cN=1$ flow geometry.
\begin{align}
C_V = \frac{\sigma_3 L^3}{G_N} \left( \frac{1}{3\epsilon^3} - \frac{3}{2\xi^2 \epsilon} + \frac{\pi}{2\xi^3} \right)
\end{align}
and 
\begin{align}
C_A = \frac{\sigma_3L^3}{8\pi^2G_N \hbar} \left( \frac{16}{3\epsilon^3} + \frac{2}{\epsilon^3}\log{L/\epsilon} - \frac{1}{2\epsilon \xi^2}(1+ 6\log{L/\epsilon}) - \frac{9\pi}{8\xi^3} \right)
\end{align}
with a cutoff $\epsilon$ in the corresponding Fefferman-Graham coordinates.\\}
%-----------------------------------------------------------------------------------
\section{Conclusions}
\label{conclusions}
In a previous study, we showed that there is a non-analytic behavior in field-theoretic complexity up to logarithmic terms in the vicinity of critical points. We demonstrated this using explicit lattice calculations and also using general scaling arguments. This term scales like $\tau^{\nu(d-1)}$ where $\nu$ is the standard critical exponent for the correlation length $\xi$ as a function of the reduced coupling $\tau$: $\xi\sim \tau^{-\nu}$. In this work, we show both, for a generic renormalization group flow geometry and also in a specific known example, that this is also true when one studies holographic complexity using the volume and action prescriptions. We also find that even though holographic complexity, like field-theoretic circuit complexity, has various ambiguities, the non-analytic term is free from such ambiguities.

\section{Acknowledgements}
 We are very grateful to Chen-Lung Hung, Sergei Khlebnikov, Nima Lashkari and Qi Zhou for comments and discussions.  In addition, we are very grateful to the DOE that supported in part this work through grant DE-SC0007884 and the DOE QuantISED program of the theory consortium ``Intersections of QIS and Theoretical Particle Physics'' at Fermilab, as well as to the Keck Foundation that also provided partial support for this work.

%----------------------------------------------------------------------------------	-----------------------------------------------------------------------	-----------------------

\appendix
\renewcommand{\theequation}{\thesection.\arabic{equation}}

\section{A Power Series solution near the boundary}
\label{power series solution}
Here we give an example of a power series solution near the boundary similar to the form used in the eq. $\ref{powerseriesf}$ for a particular choice of the potential.
We get two independent equations involving the functions $\Phi = \Phi(z)$ and $f(z)$ from eq. $\ref{equations of motion}$
\begin{align}
&\frac{(d-1)f(z)}{zf'(z)} = (\Phi'(z))^2   \\
&z^2 f(z) \left( \Phi''(z) + \frac{1}{2}\Phi'(z)\frac{f'(z)}{f(z)} - \frac{d-1}{z} \Phi'(z) \right)  -   L^2 \frac{\delta V(\Phi)}{\delta \Phi} = 0
\end{align}
We look at power series solutions when $z \ll \xi$ of the form
\begin{align}
\Phi(z) &= \tau z^{\alpha} + \sum_{k=2} \phi_k (\tau z^{\alpha})^k   \\
f(z) &= 1 + \sum_{k=1} f_k (\tau z^{\alpha})^k
\end{align}
with $\tau \sim \xi^{-\alpha}$ and $\alpha = d-\Delta$ for source deformations. Then, for the choice of the potential 
\begin{equation}
V(\Phi) = \frac{-d(d-1)}{L^2} + \frac{1}{2} m^2 \Phi^2 + \frac{1}{6L^2} g_3 \Phi^3
\end{equation}
the coefficients $\phi_k$ and $f_k$ for some small k are
\begin{align}
f_1 &= 0   \\ \non
f_2 &= \frac{\alpha}{2(d-1)}  \\   \non
\phi_2 &= \frac{- \kappa}{2\alpha (d - 3\alpha)}   \non \\
f_3 &= \frac{-2\kappa}{3\alpha(d-3\alpha)(d-1)}   \non \\
\phi_3 &= \frac{\alpha(2\alpha - d)}{4(d-1)(d-4\alpha)} + \frac{\kappa^2}{4\alpha^2(d-4\alpha)(d-3\alpha)}  \non \\
& \dots \non
\end{align}

\section{Volumes in some toy models of RG Flows}
\label{toy model}
In this appendix, we consider some toy examples of deformations by postulating functions $f(z)$ in the metric $(2.2)$. Consider the class of metrics 
\begin{align}
f_{\alpha} (z) &= (1+ (z/\xi)^{\gamma})^{\alpha} 
\end{align}
where $\alpha > 0$ is a parameter and $\gamma$ is given by the dimension of the relevant deformation from the critical point for source deformations or by the dimension of the operator getting a vev in the vev deformation. $\xi$ is related to the coupling in the case of the source deformation via the exponent $\nu$ or it is related to the vev of an operator in the vev case. These choices have the correct asymptotic form and, we explore the co-dimension one volume in these cases. The $\cN=1$ flow is a specific case in this class with $\alpha = 2$ and $\gamma = 2$ in $d=4$.\\
\commentout{The cutoffs differ by a term of $O(\epsilon^{\gamma+1})$ this time.
\begin{align}
z_0(\epsilon) - \epsilon = \frac{\alpha}{2\gamma \xi^{\gamma}}\epsilon^{\gamma+1} + O(\epsilon^{2\gamma+1})
\end{align}}
For the $d=3$ case with $\alpha = 2$, 
\begin{align}
V (\xi) &=\sigma_2 L^3 \int_{z_0(\epsilon)}^{Z_0} \frac{dz}{z^3}\frac{1}{1+z^{\gamma}/\xi^{\gamma}}  \non \\
&=\frac{\sigma_2 L^3}{(\gamma+2)z_0^2}\left(\frac{\xi}{z_0}\right)^{\gamma}  {}_2F_1[ 1, 1 + \frac{2}{\gamma} ,2 + \frac{2}{\gamma},-\frac{\xi^{\gamma}}{z_0^{\gamma}} ]
\end{align}
Using a mathematical identity that is true for $\gamma = 2$, we find that
\begin{align}
V (\xi) &= \sigma_2 L^3 \left(\frac{1}{2 z_0^2} - \frac{1}{2\xi^2}\log[1+\frac{\xi^2}{z_0^2}]\right)   \non 
\end{align}
$\gamma = 2$ holds for $\Delta = 2$ source deformations and $\Delta = 1$ vev deformations in $d=3$. 
In fact when we generalize $\alpha$ to be arbitrary but positive, for $d \geq 2$ one can analytically compute the quantity $V(\xi)$. It is given by
\begin{align}
\frac{V(\xi)}{\sigma_{d-1}L^d} =   \frac{z_0^{1-d-\alpha \gamma /2} \xi^{\alpha \gamma/2}}{(d-1+\alpha \gamma/2)} {}_2F_1[ \frac{\alpha}{2},\frac{\alpha}{2}+\frac{d-1}{\gamma},1+\frac{\alpha}{2}+
\frac{d-1}{\gamma},-\xi^{\gamma}/z_0^{\gamma} ]   
\end{align}
Using the identities
\begin{align}
{}_2F_1[ a, b; c; z ] &=   (1-z)^{-a} {}_2F_1[ a, c - b; c; \frac{z}{z-1} ]  \\
{}_2F_1[ a, b; c; z ] &= \frac{\Gamma(c) \Gamma(c-a-b)}{\Gamma(c-a)\Gamma(c-b)} {}_2F_1[ a, b; a+b+1-c; 1-z ]  \\
&+ \frac{\Gamma(c)\Gamma(a+b-c)}{\Gamma(a)\Gamma(b)}(1-z)^{c-a-b}  {}_2F_1[ c - a, c - b; 1+c - a - b; 1-z ]   \non  \\
{}_2F_1[ a, b; a; z ] &= (1-z)^{-b}   
\end{align}
the above expression can be expanded in a series of the form 
\begin{align}
\frac{V(\xi)}{\sigma_{d-1}L^d}  &= \frac{1}{(d-1)z_0^{d-1}} - \frac{\alpha}{(d-1-\gamma)z_0^{d-1-\gamma}\xi^{\gamma}} + \cO(1/z_0^{d-1-2\gamma}\xi^{2\gamma}) \\
&+ \frac{\Gamma{(\frac{\alpha}{2} + \frac{d-1}{\gamma})} \Gamma{((1-d)/\gamma)}}{\gamma \Gamma{(\alpha/2)}\xi^{d-1}}   \non
\end{align}
Thus we find a series of terms in powers of $\xi^{-\gamma}$ or equivalently $\tau^2$ and a universal term going like $\xi^{-(d-1)}$ or $\tau^{\nu(d-1)}$. The first series is analytic in the coupling $\tau$ while the universal term is non-analytic in $\tau$ in general. This matches our expectation from the general results of sec. \ref{non-analytic terms}. The exact form of $v_0$ for this class of metrics is then
\begin{align}
v_0 = \frac{\Gamma{(\frac{\alpha}{2} + \frac{d-1}{\gamma})} \Gamma{((1-d)/\gamma)}}{\gamma \Gamma{(\alpha/2)}} 
\end{align}

\commentout{
A crossover in the leading term happens in this case as
\begin{align}
\delta V(\xi) =
 \begin{cases}
 \frac{\alpha}{2\xi^{d-1}} \log(\xi/\epsilon) , \text{if} \  \gamma = d-1  \\
 \\
 \frac{\alpha (d-1)}{2\gamma(d-1-\gamma)} \frac{1}{\epsilon^{d-1-\gamma} \xi^{\gamma}}  , \text{if} \ \gamma < d-1 \\
 \\
 -\frac{2\Gamma(\frac{\alpha}{2}+\frac{d-1}{\gamma})\Gamma(\frac{1-d}{\gamma})}{\Gamma(\alpha/2)} \frac{1}{\xi^{d-1}}, \text{if}\  \gamma > d-1  \\
 \end{cases}
 \end{align}
The crossover in the leading term happens at $\gamma = d-1$ for any $\alpha > 0$. Thus we find that in this toy model, there is no universal scaling behavior in complexity but in fact, the form of the leading behavior near the critical point depends on $\Delta$. Moreover, this scaling exponent is insensitive to $\alpha$ which was a label for choosing one specific toy model. \\  }

For toy models that have AdS domain walls i.e a flow from one AdS space to another, we look at the class of metrics with 
\begin{align}
f_{\alpha}(z) = 1 + \frac{\alpha (z/\xi)^{\gamma}}{1+ (z/\xi)^{\gamma}}   
\end{align}
Here $\alpha$ is related to the $L_{IR}$ by $\alpha = L^2/L_{IR}^2 - 1$ and $\gamma$ is related to the operator dimension of the UV deformation. This can be seen by expanding $f$ close to the boundary $z << \xi$. We analytically compute the volumes in the special case when $\alpha \rightarrow 0$ and find that 
\begin{align}
\frac{V(\xi)}{\sigma_{d-1}L^d} &= \frac{1}{(d-1)z_0^{d-1}} + \frac{\alpha}{2\xi^{d-1}} \left( \frac{\pi}{\gamma} \csc{\pi (1-d)/\gamma} \right. \non \\
&+ \left. \frac{z_0^{\gamma+(1-d)}}{(\gamma+1-d)(\xi^{\gamma}+z_0^{\gamma})^{1+\frac{1-d}{\gamma}}} {}_2F_1[1+\frac{1-d}{\gamma}, 1+\frac{1-d}{\gamma}, 2+\frac{1-d}{\gamma},  \frac{z_0^{\gamma}}{\xi^{\gamma} + z_0^{\gamma}}]    \right)
\end{align}
which again gives a series of terms analytic in the coupling $\tau$ and a non-analytic term of the form $\tau^{\nu(d-1)}$.
\begin{align}
\frac{V(\xi)}{\sigma_{d-1}L^d} &= \frac{1}{(d-1)z_0^{d-1}} - \frac{\alpha}{2(d - 1 - \gamma)} \left(  \frac{1}{z_0^{d-1-\gamma} \xi^{\gamma}} - \frac{(\gamma + 1 - d)}{(2\gamma + 1 - d)}\frac{1}{z_0^{d-1-2\gamma} \xi^{2\gamma}}    \right.  \\
+& \left. \cO(\frac{1}{z_0^{d-1-3\gamma}\xi^{3\gamma}})  +  \frac{\pi \alpha}{2 \gamma \xi^{d-1}} \csc{\frac{\pi(1-d)}{\gamma}}  \right)   \non
\end{align}
We thus have for this class of toy models,
\begin{align}
v_0 = \frac{\pi \alpha}{2 \gamma} \csc{\frac{\pi(1-d)}{\gamma}} 
\end{align}
	
\commentout{
In this case, we have 
\begin{align}
\frac{\delta V(\xi)}{\sigma_2L^3} =\frac{1}{\xi^2}\log{\xi/\epsilon} + \frac{1}{2\xi^2} +  ...  
\end{align}
where the ... represents terms that go to zero as $\epsilon \rightarrow 0$.
For general $\gamma$, there is a crossover between the behavior of the leading piece as $\gamma$ crosses 2. The two terms involved in the crossover are 
\begin{align}
\frac{\delta V(\xi)}{\sigma_2L^3} = \frac{1}{\gamma}\left( \frac{2}{2-\gamma}\frac{\epsilon^{\gamma-2}}{\xi^{\gamma}} + \frac{\pi}{\xi^2} \csc 2\pi / \gamma + ....  \right)
\end{align}
\begin{align}
\delta V(\xi) =
 \begin{cases}
 \frac{1}{\xi^2} \log(\xi/\epsilon) , \text{if} \  \gamma = 2  \\
 \\
 \frac{2}{(2-\gamma)\gamma}\frac{\epsilon^{\gamma-2}}{\xi^{\gamma}}   , \text{if} \ \gamma < 2 \\
 \\
 \frac{\pi}{\gamma \xi^2} \csc 2\pi / \gamma, \text{if}\  \gamma > 2  \\
 \end{cases}
 \end{align}
The leading behaviour has an $\epsilon$ dependence when $\gamma < 2$ but is independent of $\epsilon$ when $\gamma > 2$. We get $\log$ like leading terms only at special values of $\Delta$. In every case, $\delta V(\xi)$ is positive simply because the volume of a slice in the AdS 'box' is the largest at the critical point.}

\section{Non-analytic piece as a bulk term in the action}
\label{non-analytic boundary calculation}
In this appendix, we show that if there is any universal term in the WDW action, then it necessarily comes from the bulk piece. We do this by computing the boundary action terms for a general $f(z)$ and showing that they are analytic in $\tau$. Using the affine parametrization for the null sheets, we find that $I_{null}=0$. We look at a more general parametrization for the specific example of the $\cN=1$ flow in Section \ref{ambiguity}. $I_{GHY}$ and $I_{J}$ are both evaluated on the cutoff surface $z=z_0$ and we study these for small $z_0$. \\
We find that 
\begin{align}
I_{GHY} = \frac{d \sigma_{d-1} L^{d-1}\sqrt{f(z_0)}}{4 \pi G_N z_0^{d}} \int_{0}^{z_0} \frac{dz}{\sqrt{f(z)}}
\end{align}
We see that using the near-boundary expansion for $f(z)$ in \ref{powerseriesf}, the above expression only contains integer powers of $\tau$ and hence is analytic in the coupling.\\
For $I_{J}$, we find that it is independent of $f(z)$ with no $\xi$ dependence.
\begin{equation}
I_J = \frac{\sigma_{d-1}L^{d-1}}{4 \pi G_N z_0^{d-1}} \log{L/z_0}
\end{equation}
In Section $\ref{ambiguity}$, we compare different renormalization geometries along the RG flow and look at differences in the action $\delta I$ for two geometries. The cutoff $z_0$ picks up an $\xi$ dependence because the cutoff is kept the same in Fefferman-Graham coordinates for the different geometries. In such cases, we still find that the quantities $\delta I_{GHY}$ and $\delta I_{J}$ are analytic in the reduced coupling $\tau$. 

\bibliographystyle{unsrt}
\bibliography{references}

\end{document}